\documentclass[conference, onecolumn, draft]{IEEEtran}
\IEEEoverridecommandlockouts
\usepackage{ifpdf}
\usepackage{cite}
\usepackage{graphicx}
\usepackage{amsmath}
\usepackage{algorithm}
\usepackage{algpseudocode}
\usepackage{array}
\usepackage{blindtext}
\usepackage{epstopdf}
\usepackage{bm}
\usepackage{enumerate}
\usepackage{amsbsy}
\usepackage{amssymb}
\usepackage{amsthm}
\usepackage{amscd}
\usepackage{subfigure}
\usepackage{color}
\usepackage{booktabs}
\usepackage{multirow}
\usepackage{hhline}
\usepackage{makecell}
\usepackage{bbold}
\usepackage{url}
\usepackage{tabularx}
\usepackage{cellspace}
\usepackage{boldline}
\usepackage{balance}
\usepackage[inline]{enumitem}

\def \ba           {\boldsymbol{a}}

\def \bp           {\boldsymbol{p}}

\def \bs           {\boldsymbol{s}}

\def \bx           {\boldsymbol{x}}
\def \by           {\boldsymbol{y}}

\def \bR           {\boldsymbol{R}}

\def \bzero        {\boldsymbol{0}}
\def \bone         {\boldsymbol{1}}

\def \binaryB      {\mathbb{B}}
\def \complexC     {\mathbb{C}}
\def \expecE       {\mathbb{E}}

\def \realRp       {\mathbb{R}_{+}}

\def \arg          {\mathrm{arg}}

\def \st           {\mathrm{s.t.}}
\def \ow           {\mathrm{otherwise}}

\def \degree       {^\circ}

\title{Joint Optimization of Waveform Covariance Matrix and Antenna Selection for MIMO Radar}

\author{\IEEEauthorblockN{Arindam~Bose\IEEEauthorrefmark{1}\IEEEauthorrefmark{4},
		Shahin~Khobahi\IEEEauthorrefmark{2}\IEEEauthorrefmark{4} and
		Mojtaba~Soltanalian\IEEEauthorrefmark{3}}
	\thanks{%
		\IEEEauthorrefmark{1}Corresponding author (e-mail: \textit{abose4@uic.edu}).
	} 
	\thanks{%
		\IEEEauthorrefmark{4}The first two authors contributed equally to this work.
	} 
	\thanks{%
		This work was supported in part by U.S. National Science Foundation Grants CCF-1704401 and ECCS-1809225.
	}
	\IEEEauthorblockA{Department of Electrical and Computer Engineering, University of Illinois at Chicago, Chicago, Illinois 60607, USA\\ Email: \{\IEEEauthorrefmark{1}abose4, \IEEEauthorrefmark{2}skhoba2, \IEEEauthorrefmark{3}msol\}@uic.edu}}

\begin{document}

\maketitle

\begin{abstract}
	In this paper, we investigate the problem of jointly optimizing the waveform covariance matrix and the antenna position vector for multiple-input-multiple-output (MIMO) radar systems to approximate a desired transmit beampattern as well as to minimize the  cross-correlation of the received signals reflected back from the targets.
	We formulate the problem as a non-convex program and then propose a cyclic optimization approach to efficiently tackle the problem.
	We further propose a novel local optimization framework in order to efficiently design the corresponding antenna positions.
	Our numerical investigations demonstrate a good performance both in terms of accuracy and computational complexity, making the proposed framework a good candidate for real-time radar signal processing applications.
\end{abstract}


\section{Introduction}
	Multiple-input-multiple-output (MIMO) radar has been an emerging technology during last two decades, attracting a great deal of interest from researchers in radar and signal processing communities \cite{1316398, 1399142, 1399141, 4516997, 4350230, 4358016, li2008mimo, 8683876, 4408448, 4524058, 4567663, 7811203, SOLTANALIAN2014132}. One of the main advantages of MIMO radar systems compared with the traditional phased-array radars is their ability to transmit multiple probing waveforms allowing for transmitting arbitrary waveforms (spatial diversity). Briefly speaking, the waveform diversity provided by a MIMO system can increase the resolution and sensitivity to target movements, and specifically, paving the way for applying adaptive array processing techniques. An important task in MIMO radar systems is thus to design the probing waveforms to approximate a desired beampattern, and to further minimize the cross-correlation of the signals reflected from various targets, and from reflections of other waveforms. Alternatively, one can consider the design of the probing signal covariance matrix as it provides more degrees of freedom compared to designing the waveforms directly \cite{5962371, 7915123, 6698378, 7955071, 7027831, 8378710, 7944229, 8859282, 7071108}.
	
	
	A large part of the existing research on covariance waveform design focuses mainly on the scenario with a uniform linear array (ULA) and half-wavelength inter-element spacing in order to match a desired beampattern. However, such designs are typically concerned with statistical properties of the transmitted waveforms rather than incorporating a design of the positions of the transmit antennas as well. Recently, it was shown in \cite{8378710}\nocite{8645383} that unlike a ULA configuration where the total number of antennas and their positions are fixed, one can achieve additional degrees of freedom by carefully designing the antenna positions on a grid point for approximating the transmit beampattern with the same number of antennas (distributed non-uniformly on a grid point). As a result, assuming the total number of transmit antennas is fixed, a joint optimization of the covariance matrix and the antenna selection vector can achieve superior results compared with methods operating on a ULA configuration.
	
	In this paper, we propose a novel cyclic optimization approach to efficiently tackle the non-convex nature of the joint optimization of the waveform covariance matrix and antenna positions, and furthermore, in order to efficiently design the corresponding antenna positions, we introduce a binary local optimization algorithm. Our method allows for generating waveform covariance matrices with low cross-correlation properties by exploiting the additional degree of freedom in designing the antenna positions.
	
	\begin{figure}
		\centering
		\includegraphics[draft=false, width=0.6\textwidth]{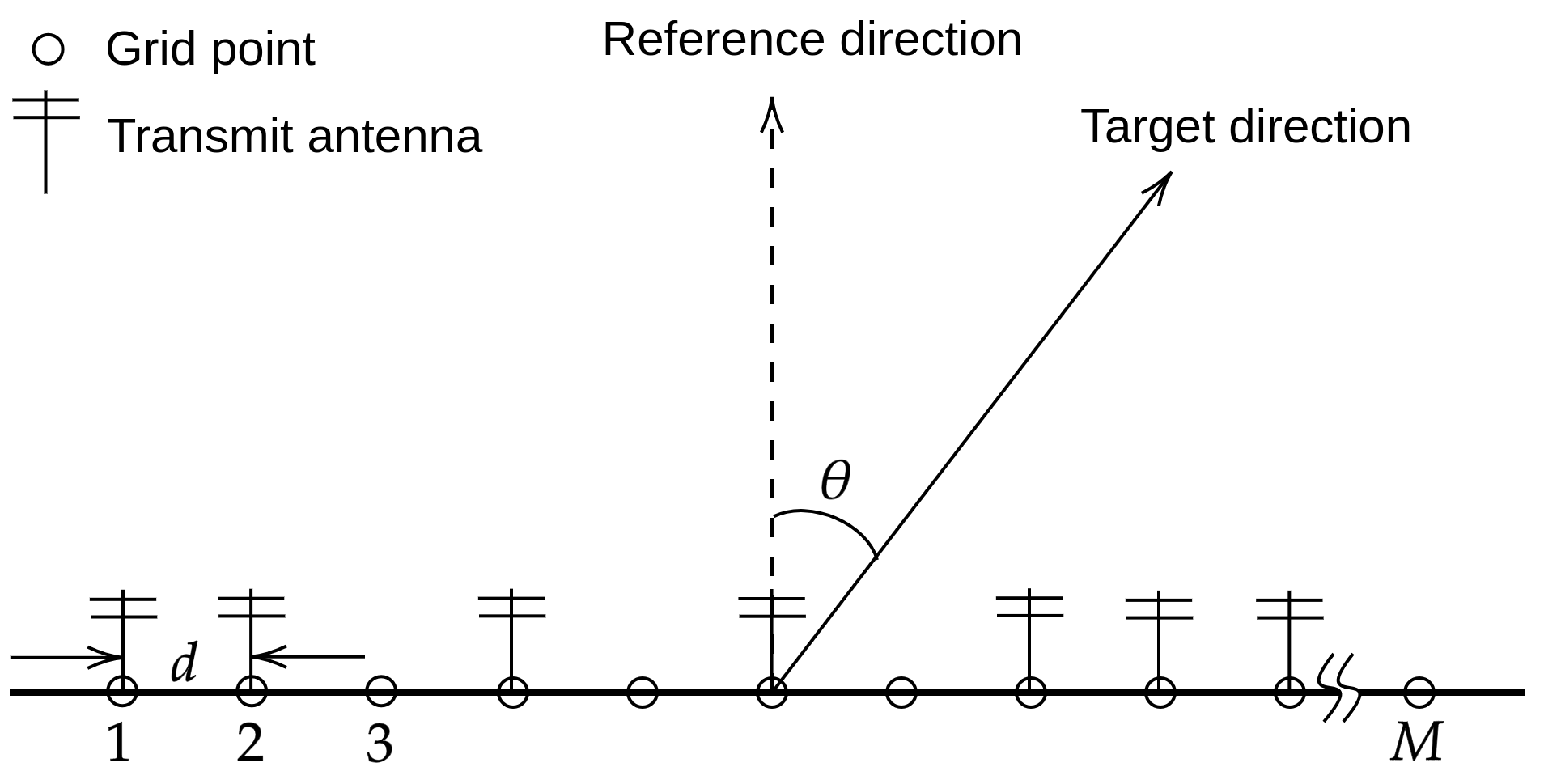}
		\caption{Geometry of a MIMO radar with $M$ grid points. Only $N$ grid points can be used for antenna placement.}
		\label{fig:radar_scema}
		\vspace{-4.5mm}
	\end{figure}

\section{Signal Model and Problem Formulation}
	We consider the problem of placing $N$ transmit antennas placed on a non-uniform linear array (ULA) positions with $M(\geq N)$ grid points with equal grid spacing $d$, in order to produce a desired beampattern as depicted in Fig. \ref{fig:radar_scema}. Let $s_m(l),$ with $m \in \{1, \cdots, M\}$ and $l \in \{1, \cdots, L\}$, denote the transmit signal from $m$-th antenna, where $L$ is the signal length in discrete-time and $\bs(l) = [s_1(l), s_2(l), \cdots, s_M(l)]^T$ is the space-time transmit waveform with length $ML$, where $(\cdot)^T$ represents the transpose of a vector/matrix. Assuming a narrow-band signal model and non-dispersive propagation, the $M$-dimensional \textit{steering vector} at an arbitrary angle $\theta$ is given by $
	\ba(\theta) = [1, e^{j\frac{2\pi}{\lambda} d\sin\theta}, \cdots, e^{j\frac{2\pi}{\lambda} (M-1)d\sin\theta}]^T $, where $\lambda$ is the wavelength of the transmitted signal. 
	
	Let us introduce a binary antenna position vector to represent the antenna configuration as
	\begin{align}
	\bp = [p_1, p_2, \cdots, p_m, \cdots, p_M]^T,~~p_m \in \{0,1\},
	\end{align}
	where $p_m=1$ indicates that the $m$-th grid point is selected for antenna placement; otherwise we have $p_m=0$. The corresponding waveform at the target location at the direction $\theta$ with respect to (w.r.t.) the ULA is then given by,
	\begin{align}
	x(l) = (\bp \odot \ba(\theta))^H\bs(l), \qquad l \in \{1,\cdots,L\},
	\end{align}
	where $\odot$ denotes the Hadamard product and $(\cdot)^H$ represents the conjugate transpose of the argument vector/matrix.
	Consequently, the power produced by the waveforms at a generic direction $\theta$ can be written as
	\begin{align} \label{eq:power}
	P(\theta) &= \expecE\{|x(l)|^2\} \\
	&= (\bp \odot \ba(\theta))^H \expecE\{\bs(l)\bs^H(l)\} (\bp \odot \ba(\theta)) \nonumber \\
	&= \bp^T\Re\left\lbrace \bR \odot \left(\ba(\theta) \ba^H(\theta)\right)^*\right\rbrace\bp \nonumber,
	\end{align}
	where
	\begin{align}  
	\bR = \expecE\left\lbrace\bs(l)\bs^H(l)\right\rbrace
	\end{align}
	is the covariance matrix of the transmit waveforms $\bs(l)$, to be designed. 
	Here $\expecE\{\cdot\}$ and $\Re\{\cdot\}$ represent the expected value and the real part of their argument, respectively.
	Furthermore, $(\cdot)^*$ denotes the conjugate of the argument vector/matrix.
	
	Let $d(\theta)$ denote the desired transmit beam-pattern, and $\{\theta_k\}_{k=1}^{K}$ be a grid of points that covers the radial sectors of interest.
	We assume that the said grid comprises of points which are good approximations of the locations of $\tilde{K}$ targets of interest that we wish to probe at locations $\{\tilde{\theta}_k\}_{k=1}^{\tilde{K}}$.
	In addition, we assume that some partial information regarding the target positions are available at hand, e.g., we possess some initial estimates $\{\hat{\theta}_k\}_{k=1}^{\hat{K}}$ of $\{\tilde{\theta}_k\}_{k=1}^{\tilde{K}}$.
	Thus one can form the desired beam-pattern as follows (with $\hat{K}$ being the resulting estimate of $\tilde{K}$):
	\begin{align*}
	d(\theta) = \left\lbrace
	\begin{array}{ll}
	1, & \theta \in [\hat{\theta}_k-\frac{\triangle}{2},                \hat{\theta}_k+\frac{\triangle}{2}],~~ k \in \{1,\cdots,\hat{K}\}, \\
	0, & \ow,
	\end{array}
	\right.
	\end{align*}
	where $\triangle$ is the chosen beam-width for each target.
	
	Our goal is to design the waveform covariance matrix $\bR$ as well as designing the antenna positions (i.e. optimizing $\bp$) such that the transmitted beampattern $P(\theta)$ approximates a given beampattern $d(\theta)$ over the radial sectors of interest in a least squares (LS) sense, and also such that the cross-correlation of the reflected waveform from the targets is minimized. One can formulate this problem by defining a cost function as follows \cite{4350230}:
	\begin{align} \label{eq:J}
		&J(\bp, \bR, \alpha) \\
		&~~ = \underbrace{\frac{1}{K}\sum_{k=1}^{K}{w_k\left|\bp^T\Re\left\lbrace \bR \odot \left(\ba(\theta_k) \ba^H(\theta_k)\right)^*\right\rbrace\bp - \alpha d(\theta_k)\right|^2}}_{\text{beampattern matching term}} \nonumber\\
		&~~ + \underbrace{\frac{2\omega_c}{\hat{K}(\hat{K} - 1)} \sum_{p=1}^{\hat{K}-1}{\sum_{q=p+1}^{\hat{K}} {\left| \bp^T\Re\left\lbrace \bR \odot \left(\ba(\hat{\theta}_p) \ba^H(\hat{\theta}_q)\right)^*\right\rbrace\bp \right|^2}}}_{\text{cross-correlation term}}\nonumber
	\end{align}
	where $ \omega_k \geq 0, k = 1,\cdots,K$, is the weight for the $k$-th radial sector and $ \omega_c \geq 0 $ is the weight for the cross-correlation term and $\alpha>0$ is a scaling parameter to be designed. In the next section, we propose our optimization method allowing us to not only optimize the covariance matrix but also the antenna positions. 
	
	
	\section{Optimization Algorithm}
	The joint optimization problem of designing the transmitted waveform covariance and the antenna position can be formulated as
	\begin{subequations} \label{eq:prob}
		\begin{align}
			\min_{\bp,\bR, \alpha}\qquad &J(\bp,\bR,\alpha) \\
			\st\qquad & \bR \succeq \bzero, \label{eq:prob1}\\
			& R_{mm} = \frac{c}{M},~~\text{for}~m = 1, \cdots, M, \label{eq:prob2}\\
			& \|\bp\|_1 = N, \label{eq:prob3}\\
			& p_m = \{0,1\}, ~~~\text{for}~m = 1, \cdots, M,\label{eq:prob4}\\
			& \alpha > 0
		\end{align}
	\end{subequations}
	Since $\bR$ is a covariance matrix, it must be positive semidefinite as well as all antennas are required to transmit uniform power. These two conditions are enforced in constraints \eqref{eq:prob1} and \eqref{eq:prob2}. Furthermore, the constraints \eqref{eq:prob3} and \eqref{eq:prob4} guarantee that only $N$ antennas are to be placed in $M(>N)$ possible grid points, and that the vector $\bp$ is binary.
	
	It is not hard to verify that the optimization problem in \eqref{eq:prob} is mixed Boolean-nonconvex in nature and hard to solve for a global solution. In order to tackle such non-convexity, we propose a \emph{cyclic optimization} approach with respect to the design variables $(\bR,\alpha)$ and $\bp$. Note that, although the optimization problem w.r.t. the antenna position vector is non-convex, our approach converges to a good local minima quickly. \nocite{8314765}
	
	\subsection{Optimization for $\bR$ and $\alpha$}\label{subsec:optR} For a fixed $\bp$, the solution to the minimization problem with respect to the design variables $(\bR,\alpha)$ in the $t$-th iteration can be cast as
	\begin{align}\label{eq:optR}
		\left(\bR^{(t)},\alpha^{(t)}\right) =\arg\min_{\bR,\alpha} &~J(\bp^{(t-1)},\bR,\alpha)\\
		\st &~\bR \succeq \bzero,~\alpha > 0,\nonumber \\ & R_{mm} = \frac{c}{M},~\text{for}~m = 1, \cdots, M.\nonumber
	\end{align} 
	It is easy to verify that the above optimization problem can be reformulated as a constrained convex quadratic program, and hence, can be solved efficiently using off-the-shelf convex solvers (such as CVX \cite{cvx}). 
	
	\subsection{Optimization for $\bp$}\label{subsec:optp} For fixed $(\bR,\alpha)$, the solution to the optimization of the antenna selection vector $\bp$ can be written as follows
	\begin{align}\label{eq:optp}
		\bp^{(t+1)} = \arg\min_{\bp}&~J(\bp,\bR^{(t)},\alpha^{(t)}),\\
		\st &~\|\bp\|_1 = N, \qquad \bp\in\{0,1\}^M \nonumber,
	\end{align}
	which we solve using the following proposed local binary optimization framework. Especially, we develop an optimization approach equipped with a simple local search procedure.
	In the following, we discuss the proposed method in order to design the antenna position vector $\bp$, in detailed manner.
	
	For a given $(\bR,\alpha)$, let us denote the objective function \eqref{eq:J} as $J(\bp)$ whose solution $\bp$ is a binary vector of length $M$ with $N$ non-zero elements.
	In other words, our search space is none other than a subset of vertices of a hypercube in an $M$-dimensional space, which is discrete with bounded cardinality.
	Hence, we undertake a deterministic strategy as opposed to stochastic approaches in order to find a solution in an iterative manner.
	Note that, the binary vector $\bp$ of length $M$ represents a hypercube with $2^M$ vertices.
	Given the solution $\bp^{(k)}$ (\textit{parent solution}) at iteration $k$, a new set of \textit{candidate solutions} $\bp^{(k+1)}_{\text{CS}}$ is generated as follows:
	\begin{align}
	\bp^{(k+1)}_{\text{CS}} = \left\{\bp~|~H\left(\bp,\bp^{(k)}\right) = 1,~\|\bp\|_1<\|\bp^{(k)}\|_1   \right\},
	\end{align}
	where $H(\bx,\by)$ denotes the Hamming distance between the two vectors, and is defined to be the number of positions $i$ such that $x_i \neq y_i$, where the subscript $i$ denotes the $i$-th element of the corresponding vector.
	In other words, given a parent solution $\bp^{(k)}$, the new set of candidate solutions is generated as the set of vectors which only differs from $\bp^{(k)}$ in one bit (with one less non-zero element only).
	Hence, the cardinality of the new candidate solution is upper bounded by $\left|\bp_{\text{CS}}^{(k+1)}\right|\leq\|\bp^{(k)}\|_1$. \looseness=-1 
	
%
	
	The next task is to select and propagate the best candidate solution (i.e., the one with the lowest objective value) to the next iteration of the algorithm.
	Given the current set of candidate solutions $\bp_{\text{CS}}^{(k)}$, we select the best solution $\bp^{(k)}$ to be considered for generating new candidate solutions at the next stage as follows:
	\begin{equation}\label{eq:21}
		\bp^{(k)} = \arg\min_{\bp\in\bp_{\text{CS}}^{(k)}}{J(\bp)}.
	\end{equation}
	Next, the solution $\bp^{(k)}$ is used as the seed for generating new candidate solutions in the next iteration of the algorithm.
	Note that the above selection strategy is a one-step local search on the objective function $J(\bp)$ on a subset of vertices of a hypercube of dimension $M$.
%
	
	Let $\binaryB_N^M$ be the set of all vertices of an $M$-dimensional hypercube with $N$ non-zero elements.
	Clearly, we aim to find the optimal antenna selection vector $\bp^{*} \in \binaryB_N^M$.
	Note that, once the selection procedure selects a vector $\bp^{(k)}$ as its output such that $\bp^{(k)}\in\binaryB_N^M$ or equivalently $\|\bp^{(k)}\|_1=N$, then one can easily argue that a locally (or possibly globally) optimal solution is obtained and that $\bp^{*}=\bp^{(k)}$ for the $k$-th iteration.
	This can be seen by noting that $\bp^{(k)}\in\binaryB_N^M$ implies $\bp^{(k-1)}\in\binaryB_{N+1}^{M}$.
	Hence, one can conclude that if $\bp^{(k)}\in\binaryB_N^M$, then $\bp^{(k)}$ is a local optimal point in a 1-Hamming distance neighborhood of $\bp^{(k+1)}$ such that $\|\bp^{(k)}\|_1 < \|\bp^{(k-1)}\|_1$, and that $\bp^{(k-1)}\in\binaryB_{N+1}^{M}$.
	Moreover, the cardinality of the search space in the 1-Hamming distance local search in \eqref{eq:21} is at most $\|\bp^{(k-1)}\|_1$ (i.e., as we had earlier that $\left|\bp_{\text{CS}}^{(k)}\right|\leq\|\bp^{(k-1)}\|_1$), and as a result the search space is reduced in each (inner) iteration. \looseness=-1

	As it was discussed earlier, we consider an alternating (cyclic) optimization approach to solve the joint optimization of covariance matrix and the antenna position vector.	Finally, the proposed cyclic optimization approach is summarized in Table \ref{tb:fullalgo}.
	\begin{table}[tp]
		\footnotesize
		\caption{The Proposed Joint Optimization Method}
		\label{tb:fullalgo}
		\setlength{\extrarowheight}{5pt} \centering
		\begin{tabular}{p{3.3in}}
			\hline \hline
			\textbf{Step 0}: Initialize the antenna position vector $\bp^{(0)} = \bone_M$, the complex covariance matrix $\bR^{(0)}\in\complexC^{N\times N}$, and the scaling factor $\alpha^{(0)}\in \realRp$, and the outer loop index $t=1$.\\
			
			\textbf{Step 1}: Solve the convex program of \eqref{eq:optR} using the procedure described in Section \ref{subsec:optR} and obtain $\left(\bR^{(t)},\alpha^{(t)}\right)$.\\
			
			\textbf{Step 2}: Employ the proposed local binary optimization approach described in Section \ref{subsec:optp} and solve the antenna position design program of \eqref{eq:optp} to obtain the vector $\bp^{(t+1)}$.\\
			
			\textbf{Step 3}: Repeat steps 1 and 2 until a pre-defined stop criterion is satisfied, e.g. $H\left(\bp^{(t)},\bp^{(t-1)}\right) = 0$.\\
			\hline \hline
		\end{tabular}
	\vspace{-6mm}
	\end{table}
	
\section{Numerical Examples}
	In this section, we provide several numerical examples in order to assess the performance of our proposed algorithm. We compare our method with the ADMM-based algorithm proposed in \cite{8378710}. In the following experiments, we assume a colocated narrow-band MIMO radar with a non-uniform linear array with $ M = 15 $ grid points with half-wavelength inter-grid interval i.e., $d = \lambda/2$, unless stated otherwise, and $N = 10$ antennas. The range of angle is $ (-90\degree, 90\degree) $ with $ 1\degree $ resolution. We set the weights for the $k$-th angular direction as $ w_k = 1$, for $ k=1,\cdots, K $; and the weight of the cross-correlation term as $w_c = 1$. \looseness=-1
	
\begin{figure}
		\centering
		\includegraphics[draft=false, width=0.7\textwidth]{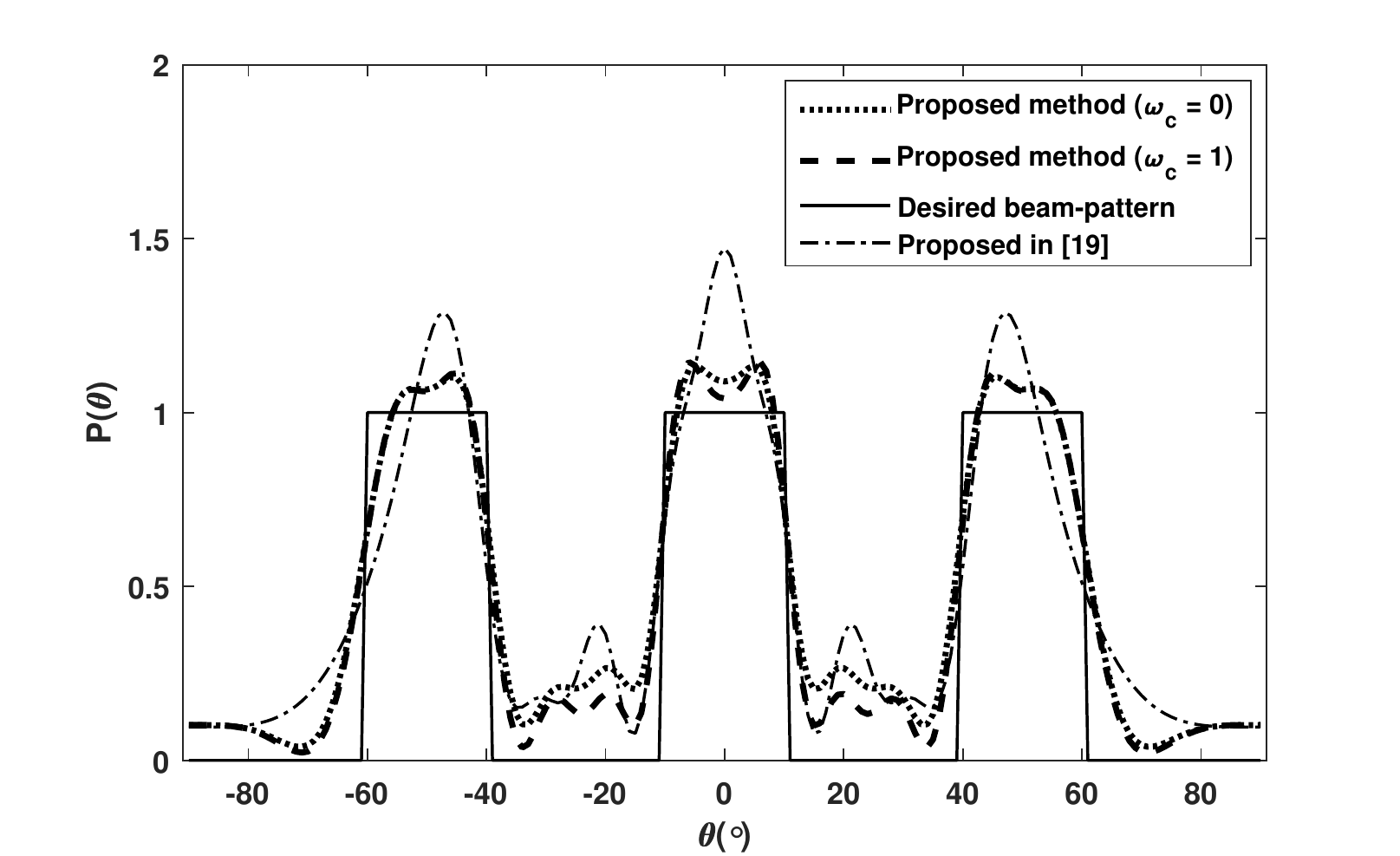}
		\caption{The transmit beampattern design for $ M = 15, N = 10 $ with and without the cross-correlation suppression with three mainlobes at $\theta = \{-50\degree, 0\degree, 50\degree\}$ with a beamwidth $20\degree$. Note that the designed beampatterns obtained with and without considering the cross-correlation term are similar to one another. However, the cross-correlation behavior of the former is much better than that of the latter in that the reflected signal waveforms corresponding to using $ \omega_c = 1 $ are almost uncorrelated with each other.} 
		\label{fig:beampattern_three_lobes}
	\end{figure}

\begin{figure}
		\centering
		\includegraphics[draft=false, width=0.7\textwidth]{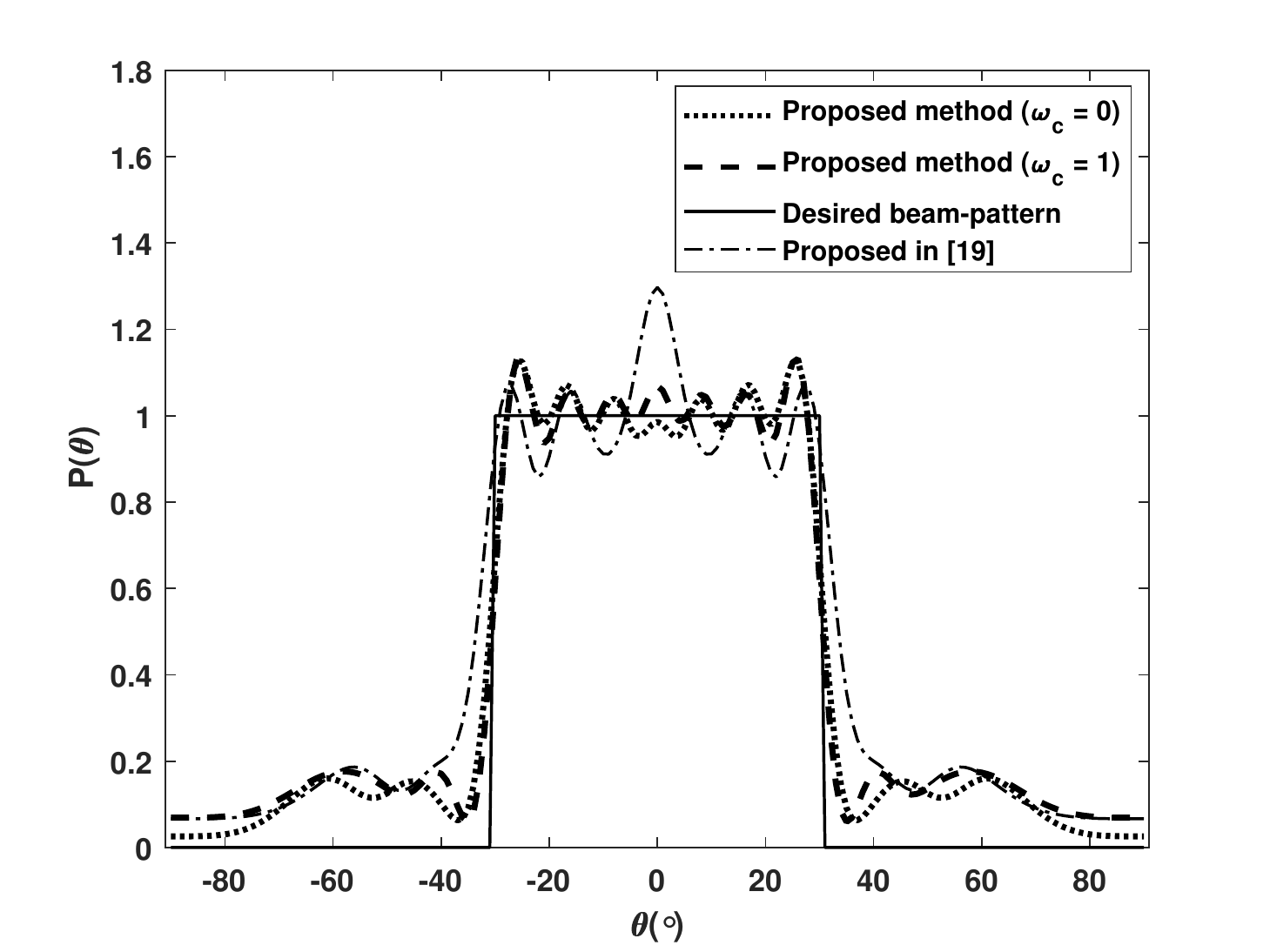}
		\caption{The transmit beampattern design for $ M = 15, N = 10 $ with and without the cross-correlation suppression with one mainlobe at $\theta = 0\degree$ with a beam-width of $60\degree$. Note that in both cases of $\omega_c=0$ and $\omega_c=1$ our proposed method can accurately approximate the desired beampattern.} 
		\label{fig:beampattern_one_lobe}
	\end{figure}

\begin{figure}
		\centering
		\includegraphics[draft=false, width=0.7\textwidth]{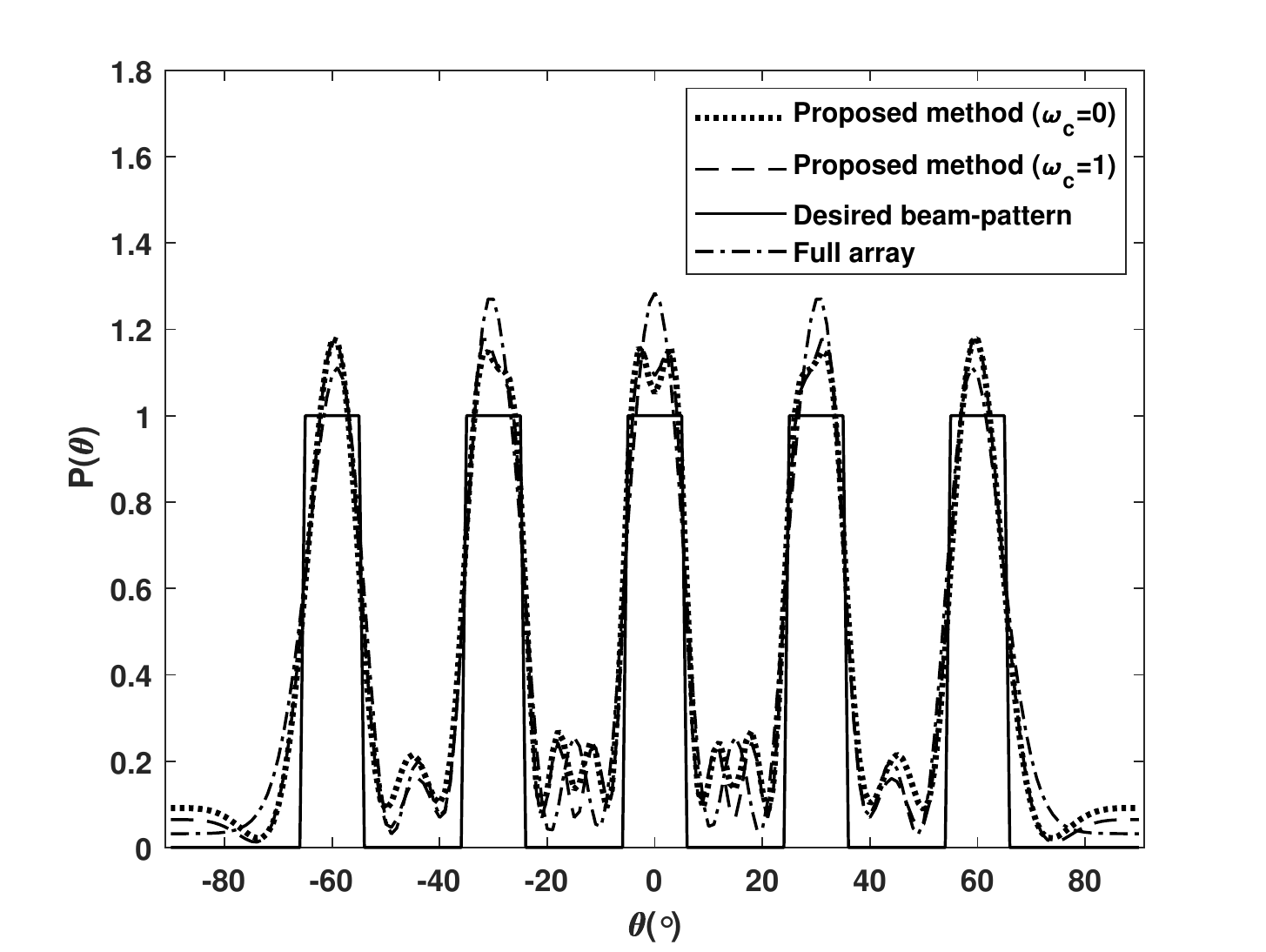}
		\caption{The transmit beampattern design for $ M = 20, N = 15 $ with and without the cross-correlation suppression with five mainlobes at $\theta = \{-60\degree, -30\degree, 0\degree, 30\degree, 60\degree\}$ with a beamwidth of $10\degree$.} 
		\label{fig:beampattern_five_lobes}
	\end{figure}

\begin{figure}
	\centering
	\includegraphics[draft=false, width=0.7\textwidth]{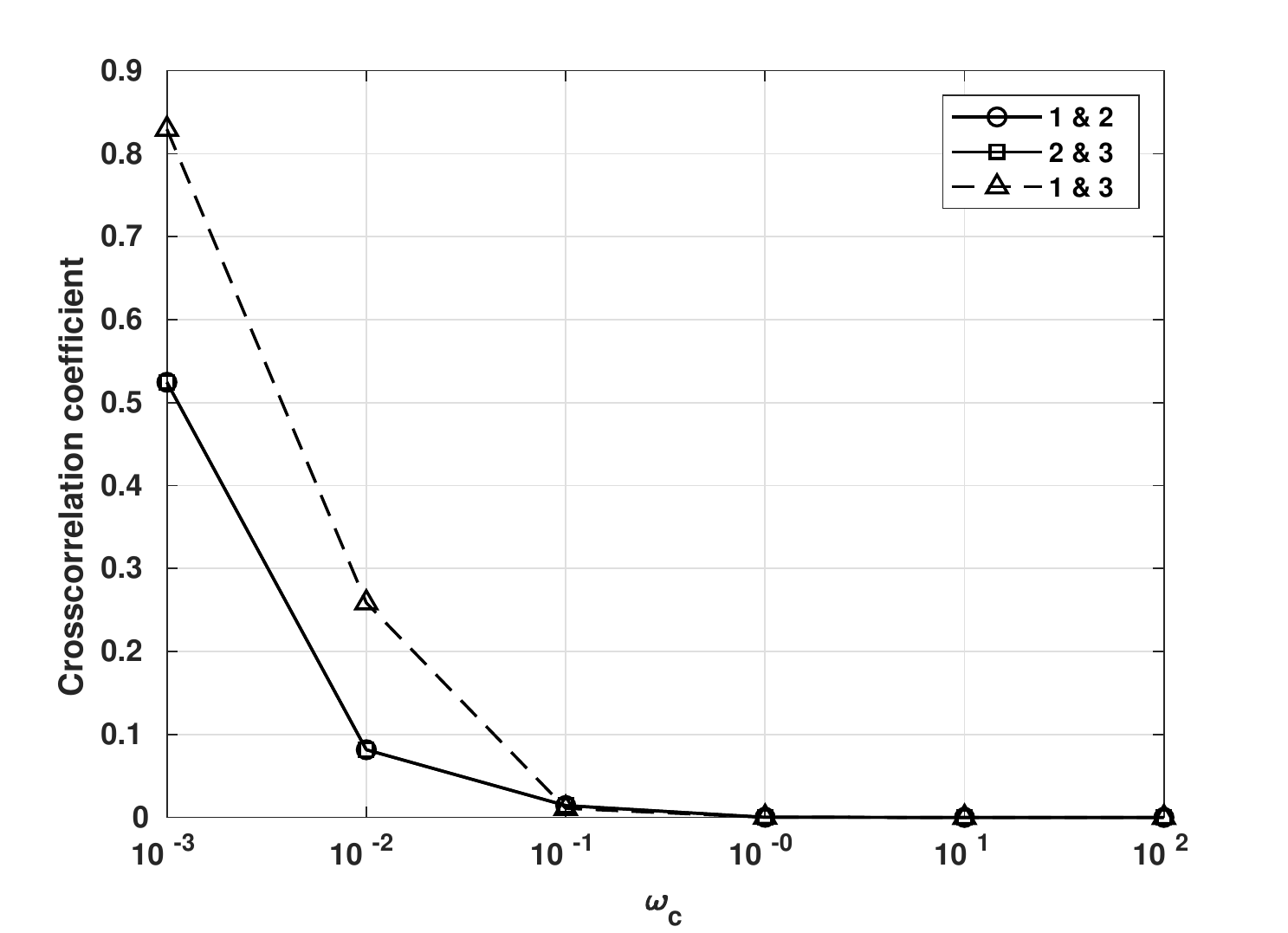}
	\caption{The comparison of the normalized magnitudes of the cross-correlation coefficients (as formulated in the second term of the right hand side of \eqref{eq:J}) for three targets of interest at directions $\theta = \{-50\degree, 0\degree, 50\degree\}$, as functions of $\omega_c$. Note that when $\omega_c$ is very small (close to zero), the first and third reflected signals are highly correlated. On the other hand, for $\omega_c>0.1$ all cross-correlation coefficients are approximately zero.} 
	\label{fig:crosscorrelation_coefficients_wc}
\end{figure}

	In Fig. \ref{fig:beampattern_three_lobes} we compare the resulting beampattern with the desired one for the two scenarios of $\omega_c=0$ and $\omega_c=1$. In addition we provide the simulation results of \cite{8378710} for three mainlobes at  $\theta = \{-50\degree, 0\degree, 50\degree\}$. In Fig. \ref{fig:beampattern_one_lobe}, we consider approximating the beampatterns with one mainlobe at $\theta =  0\degree$, and a beamwidth of $ 60\degree$. Furthermore, in Fig.~\ref{fig:beampattern_five_lobes}, we consider approximating the beampattern with $\theta = \{-60\degree, -30\degree, 0\degree, 30\degree, 60\degree\}$ and a beamwidth of $10\degree$. As it can be seen from Figs \ref{fig:beampattern_three_lobes}--\ref{fig:beampattern_five_lobes}, our proposed method can accurately match the desired beampattern. Also, note that our propose algorithm outperforms the one proposed in \cite{8378710} in terms of accuracy, and moreover, is capable of designing waveform covariance matrix with low cross-correlation, unlike \cite{8378710}. Further note that the designed beampatterns obtained with $ \omega_c = 0 $ and with $ \omega_c = 1 $ are similar to one another. However, the cross-correlation behavior of the former is much better than that of the latter in that the reflected signal waveforms corresponding to using $ \omega_c = 1 $ are almost uncorrelated with each other. This can be further verified from Fig. \ref{fig:crosscorrelation_coefficients_wc} in which we provide the comparison of the normalized magnitudes of the cross-correlation coefficients (as formulated in the second term of the right hand side of \eqref{eq:J}) for three targets of interest at directions $\theta = \{-50\degree, 0\degree, 50\degree\}$, as functions of $\omega_c$.
	
\begin{figure}
	\centering
	\includegraphics[draft=false, width=0.7\textwidth]{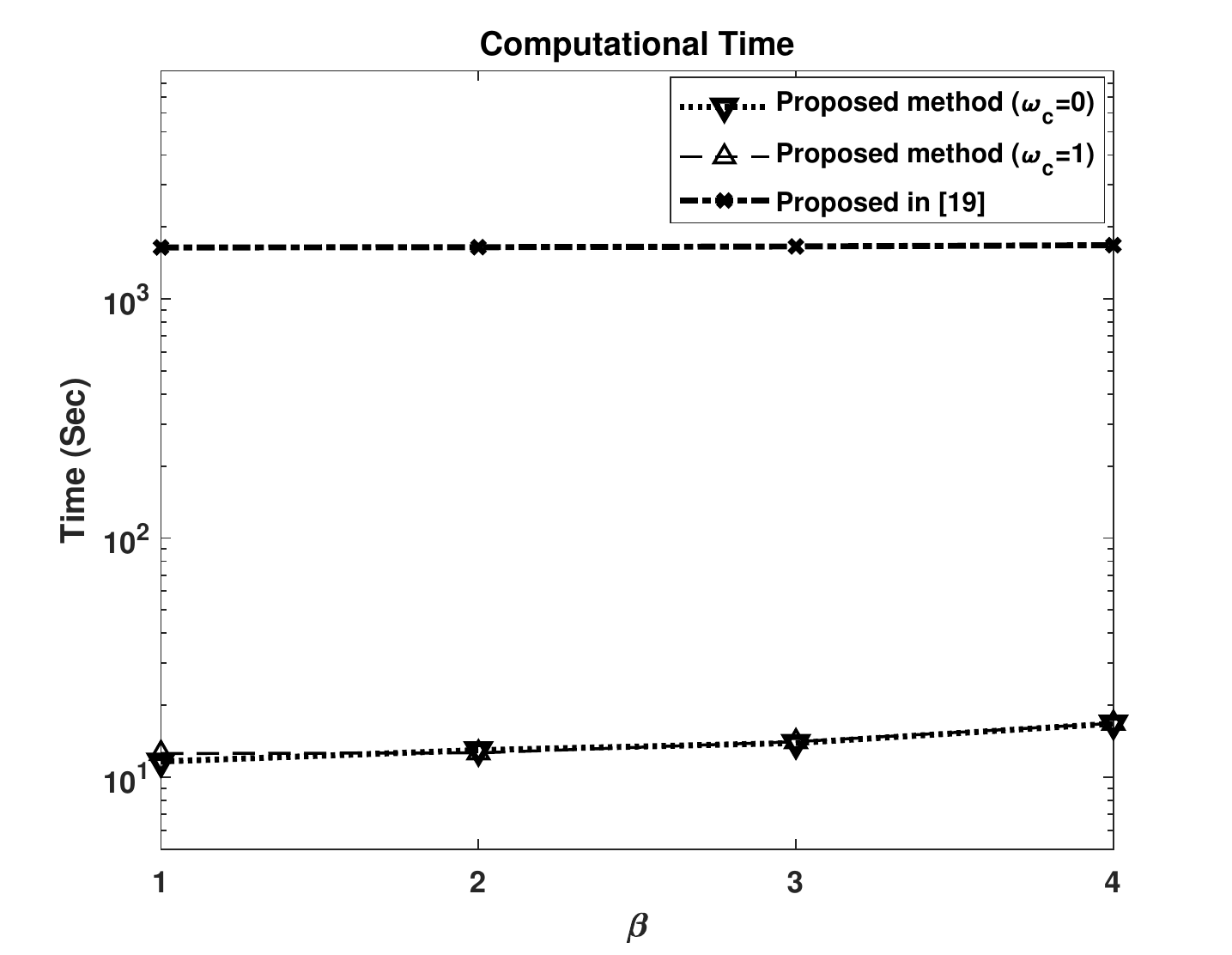}
	\caption{Comparison of the computational cost of the proposed algorithm and that of the method in \cite{8378710} for different number of grid points and that of antennas. We consider $M=4$ and $N=3$ as initialization, and then linearly scale $M$ and $N$ by the factor of $\beta \in\{1,2,3,4\}$. The proposed algorithm significantly outperforms the ADMM-based method proposed in \cite{8378710} by a factor of more than $100$, resulting our algorithm particularly suitable for real-time applications.} 
	\label{fig:runtime_final}
\end{figure}

\begin{figure}
	\centering
	\includegraphics[draft=false, width=0.75\linewidth]{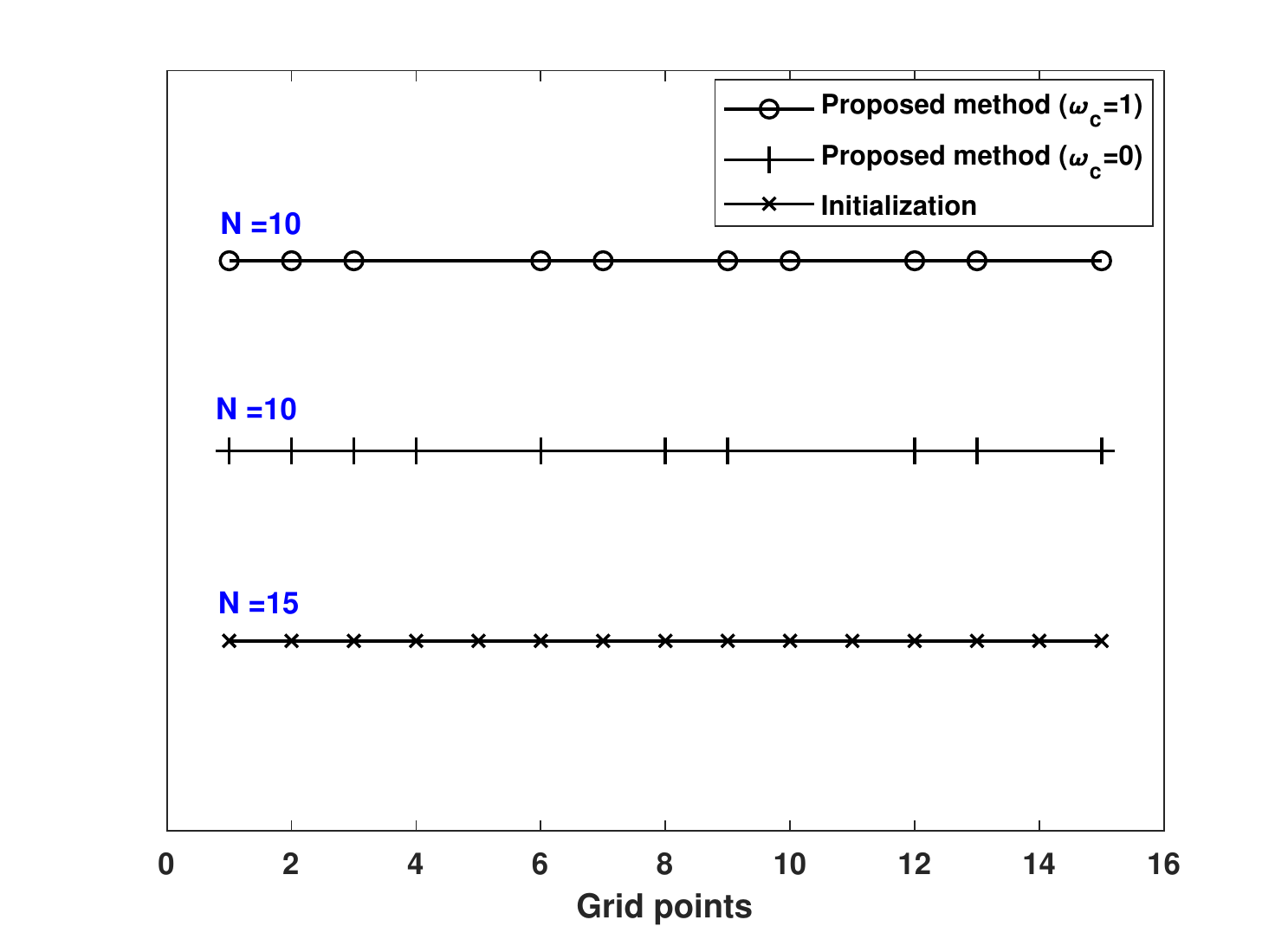}
	\caption{The antenna position $\bs$ for $ M = 15, N = 10 $ with and without the cross-correlation suppression. $y$-axis is used only for representation purposes.} 
	\label{fig:final_selected_antenna}
\end{figure}

	In Fig. \ref{fig:final_selected_antenna}, we demonstrate the final antenna position vectors suggested by the proposed algorithm for the two cases of $ \omega_c = 0 $ and $ \omega_c = 1 $. Finally, Fig. \ref{fig:runtime_final} demonstrates the computational cost of our proposed algorithm and that of proposed in \cite{8378710}. Note that our proposed algorithm significantly reduces the computational cost of the ADMM-based method in \cite{8378710} by a factor of more than $100$, making our algorithm particularly suitable for real-time applications.
	
	\section{Conclusion} \label{sec:con}
	In this paper, the problem of jointly designing the probing signal covariance matrix as well as the antenna positions to approximate a given beampattern was studied. 
	In order to tackle the problem, we proposed a novel cyclic optimization method based on the non-convex formulation of the problem. In addition, we used a local optimization algorithm to tackle the non-convex problem of designing antenna positions.
	Several numerical examples were provided which demonstrates the superiority of the proposed method over the existing ADMM-based method in terms of accuracy and computational complexity.
	

\newpage
\bibliographystyle{IEEEbib}
\balance

\bibliography{refs}

\end{document}